\documentclass[aps,reprint,notitlepage,floatfix,superscriptaddress]{revtex4-1}
\usepackage[pdftex]{graphicx}

\usepackage{amsmath,amssymb,amstext}
\usepackage{mathrsfs}

\usepackage{multirow}

\usepackage{comment}
\usepackage{bbold}
\usepackage{dsfont}
\usepackage{color}
\usepackage{hyperref}

\usepackage{tikz}

\newcommand{\ket}[1]{\left|#1\right>}

\begin{document}
\title{Reconstructing ultrafast energy-time entangled two-photon pulses}

  \author{Jean-Philippe W. MacLean}
  \email{jpmaclean@uwaterloo.ca}
  \affiliation{Institute for Quantum Computing, University of Waterloo, Waterloo,
  Ontario, Canada, N2L 3G1} 
  \affiliation{Department of Physics \& Astronomy, University of Waterloo,
  Waterloo, Ontario, Canada, N2L 3G1}
  \author{Sacha Schwarz}
  \email{sacha.schwarz@uwaterloo.ca}
  \affiliation{Institute for Quantum Computing, University of Waterloo, Waterloo,
  Ontario, Canada, N2L 3G1} 
  \affiliation{Department of Physics \& Astronomy, University of Waterloo,
  Waterloo, Ontario, Canada, N2L 3G1}
  \author{Kevin J. Resch}
  \affiliation{Institute for Quantum Computing, University of Waterloo, Waterloo,
  Ontario, Canada, N2L 3G1} 
  \affiliation{Department of Physics \& Astronomy, University of Waterloo,
  Waterloo, Ontario, Canada, N2L 3G1}

\begin{abstract}
 The generation of ultrafast laser pulses and the reconstruction of their
 electric fields is essential for many applications in modern optics.
 Quantum optical fields can also be generated on ultrafast time scales,
 however, the tools and methods available for strong laser pulses are not appropriate
 for measuring the properties of weak, possibly entangled pulses.  Here, we
 demonstrate a method to reconstruct the joint-spectral amplitude of a
 two-photon energy-time entangled state from joint measurements of the frequencies
 and arrival times of the photons, and the correlations between them. Our
 reconstruction method is based on a modified Gerchberg-Saxton algorithm. Such
 techniques are essential to measure and control the shape of ultrafast
 entangled photon pulses. 
\end{abstract}
   
  \maketitle
 
\section{Introduction}
 The generation, control, and measurement of
 high-dimensional entangled quantum states of light are important for optical
 computing and communication~\cite{lanyon_simplifying_2009, kues_-chip_2017,
 cai_multimode_2017, sparrow_simulating_2018}.  One form of this entanglement,
 in the energy-time degree of freedom, can exhibit strong correlations in
 frequency and time~\cite{van_loock_detecting_2003,shalm_three-photon_2013},
 nonlocal interference phenomena~\cite{franson_bell_1989,
 kwiat_high-visibility_1993}, and dispersion
 cancellation~\cite{franson_nonlocal_1992, wasak_entanglement-based_2010}, with
 applications in high-capacity quantum key distribution
 \cite{nunn_large-alphabet_2013, zhong_photon-efficient_2015}, enhanced
 spectroscopy~\cite{raymer_entangled_2013},
 sensing~\cite{zhang_entanglement-enhanced_2015}, and two-photon
 absorption~\cite{dayan_two_2004}. The generation and control of energy-time
 entanglement has been realized in both bulk crystals and waveguide
 structures~\cite{kang_two-photon_2014, donohue_spectrally_2016,
 donohue_quantum-limited_2018, arzani_versatile_2018, ansari_tailoring_2018},
 however, it remains an important challenge to reconstruct the quantum state of
 the photons produced. The performance of any quantum optical technology using
 time and frequency depends on being able to both shape and 
 completely characterize such photonic states.

 In ultrafast optics and laser physics, the ability to measure the amplitude
 and phase of laser pulses on ultrafast timescales is essential for nonlinear
 optics and spectroscopy.  In this context, the problem of electric field
 reconstruction has been extensively
 studied~\cite{walmsley_characterization_2009}. Optical pulses can be produced
 on time scales much shorter than any photodetector response
 time~\cite{keller_recent_2003}, and consequently, the only thing fast enough
 to measure an ultrafast laser pulse is another ultrafast pulse.  Techniques
 such as FROG~\cite{kane_single-shot_1993} and
 SPIDER~\cite{iaconis_spectral_1998} make use of nonlinear optical processes to
 measure and reconstruct ultrafast pulses. However, adapting them to quantum
 states of light is challenging due to the low power levels of single photons.
 In addition, the algorithms developed for laser pulses do not account for the
 possibility that photons can be entangled.  New innovations are therefore
 needed to reconstruct the joint state of entangled ultrafast photon pulses. 

 Approaches for characterizing the optical modes of photons have been explored
 using homodyne measurements~\cite{lvovsky_continuous-variable_2009,
 lvovsky_quantum_2001, aichele_optical_2002, polycarpou_adaptive_2012,
 morin_experimentally_2013, qin_complete_2015}, two-photon interference
 effects~\cite{chen_measuring_2015,tischler_measurement_2015,tiedau_quantum_2018},
 and two-photon absorption in
 semiconductors~\cite{boitier_photon_2011}. The increased interest
 in time-frequency modes has also led to nonlinear ultrafast approaches for
 characterization~\cite{ansari_temporal-mode_2017, ansari_tomography_2018,
 davis_measuring_2018, davis_measuring_2018-1}. To measure both the frequency
 and time intensity correlations of energy-time entangled states, optical
 methods based on optical gating and frequency resolved measurements have
 recently been developed. These have been used to observe nonlocal dispersion
 cancellation~\cite{maclean_direct_2018} and two-photon quantum
 interferometry~\cite{maclean_ultrafast_2018} on time scales inaccessible to
 standard photodetectors. For complete characterization, however, the joint
 spectral phase is also required.  This additional phase information is
 important to understand the nature of the entanglement and to control and
 optimize the performance of quantum information protocols using heralded and
 multi-photon states~\cite{zielnicki_joint_2018}. 
 
 Recovering the phase of a field from intensity measurements in Fourier-related
 domains is known as a phase-retrieval problem.  In 1972, Gerchberg and Saxton
 provided a practical solution to this problem. They introduced an iterative
 algorithm, referred to as the Gerchberg-Saxton algorithm (GS), to extract the
 complete wavefunction of an electron beam, including its phase, from intensity
 recordings in the image and diffraction
 planes~\cite{gerchberg_practical_1972}. Their algorithm can be applied to
 problems involving electromagnetic
 waves~\cite{fienup_phase_2013,shechtman_phase_2015} including optical
 wavelengths~\cite{peri_optical_1987}.

 In this paper, we implement a technique to recover the phase of ultrafast
 energy-time entangled two-photon pulses produced via spontaneous parametric
 downconversion (SPDC) and which is based on intensity measurements of the
 frequency and the arrival time of the photons. Inspired by the conventional
 phase retrieval problem, we develop an algorithm based on a method of
 alternate
 projections~\cite{gerchberg_practical_1972,fienup_reconstruction_1978,
 fienup_phase_1982} that iterates between the frequency and time domains
 imposing the measured intensity constraints at each iteration. Measurements in
 frequency are performed with single-photon spectrometers and measurements in
 time are implemented via optical gating with an ultrafast optical laser pulse. 

\begin{figure}[t!]
 \centering
\includegraphics{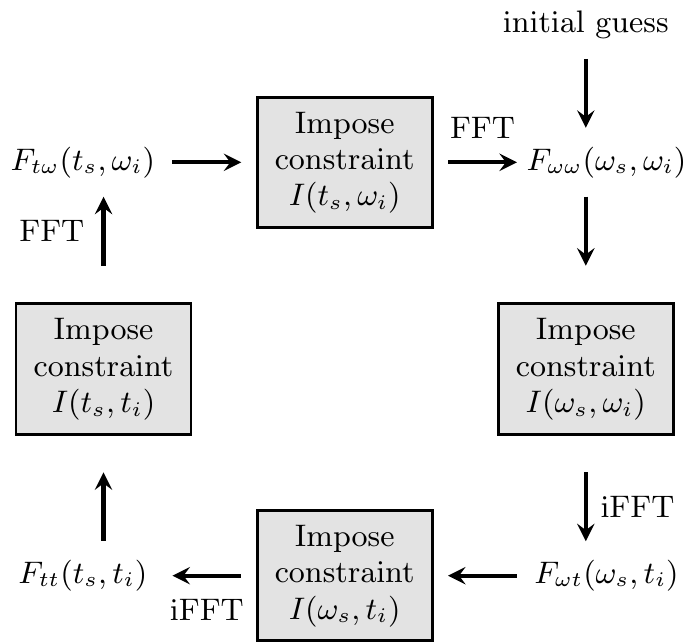}
 \caption{Block diagram of the algorithm for phase retrieval of an energy-time
   entangled two-photon state.  The algorithm is seeded with an initial guess
   of the state,
    $F_{\omega\omega}(\omega_s,\omega_i)$. 
   At every iteration, the Fast Fourier Transform is applied to transform the
   state between its frequency and time representations for both the signal and
   the idler photons. After each transformation, the magnitude of the state is
   replaced with the deconvolved measured  intensity
    data while the phase of the state is preserved.  At each
   iteration the error between the measured and recovered
   intensities either remains the same or is reduced.}
   \label{fig:mgsa}
 \end{figure}

\section{Theory}
 A pure energy-time two photon state produced via SPDC can be
 modelled as~\cite{shalm_three-photon_2013,donohue_spectrally_2016}, 
 \begin{align}
   \ket{\psi}=\int d\omega_s d\omega_i F_{\omega\omega}(\omega_s,\omega_i)
   a_s^\dagger(\omega_s)a_i^\dagger(\omega_i)\ket{0},
   \label{eq:psi}
 \end{align}
 corresponding to a superposition of frequency modes for the signal
 $a_s^\dagger(\omega_s)$ and the idler $a_i^\dagger(\omega_i)$ weighted by the
 joint spectral amplitude (JSA) function $F_{\omega\omega}(\omega_s,\omega_i)$.
 The joint spectral amplitude, 
 \begin{align}
 F_{\omega
 \omega}(\omega_s,\omega_i)=|F_{\omega\omega}(\omega_s,\omega_i)|\exp\left[i\phi(\omega_s,\omega_i)\right],
 \end{align}
 describes the amplitude, $|F_{\omega\omega}(\omega_s,\omega_i)|$, and phase,
 $\phi(\omega_s,\omega_i)$, of the state. For downconversion, it is related to
 the pump properties and the phase matching conditions of the nonlinear
 material~\cite{mosley_heralded_2008}. 
 In this form, the joint-spectral intensity
 $|F_{\omega\omega}(\omega_s,\omega_i)|^2$ characterizes
 the frequency correlations, whereas the 
 joint temporal amplitude (JTA), 
 \begin{align}
   F_{tt}(t_s,t_i)=\smallint d\omega_sd\omega_i
   F_{\omega\omega}(\omega_s,\omega_i)e^{-i\omega_st_s}e^{-i\omega_it_i},
   \label{}
 \end{align}
 which is related to the JSA by the Fourier transform, and the corresponding
 joint temporal intensity (JTI), $|F_{tt}(t_s,t_i)|^2$, characterize
 the temporal correlations. 
 Energy-time entanglement is then witnessed when the time-bandwidth product is
 found to be less than one, $\Delta(\omega_s + \omega_i)\Delta(t_s-t_i)< 1$, where
 $\Delta$ represents the standard deviation in the joint spectral and joint
 temporal intensity~\cite{mancini_uncertainty_2002,cho_uncertainty_2014,maclean_direct_2018}. 

 One is typically interested in determining the complex-valued functions
 $F_{\omega\omega}(\omega_s,\omega_i)$ or $F_{tt}(t_s,t_i)$, but only has
 access to their intensities, $|F_{\omega\omega}(\omega_s,\omega_i)|^2$ or 
 $|F_{tt}(t_s,t_i)|^2$.  The GS algorithm was originally designed to recover
 the phase from two similar intensity measurements.  However, phase retrieval
 algorithms of this form have a well-known ambiguity.  If the intensity
 distribution in the Fourier plane is centro-symmetric, then the complex
 conjugate of any given solution in the object plane is also a
 solution~\cite{guizar-sicairos_understanding_2012}.  For the energy-time
 degree of freedom, this implies a time-reversal ambiguity, i.e., it is not
 possible to distinguish between positive and negative dispersion from the
 intensity correlations in frequency and time,
 $|F_{\omega\omega}(\omega_s,\omega_i)|^2$ and $|F_{tt}(t_s,t_i)|^2$, alone. In
 the present work, we measure properties of entangled photons and can naturally
 include other time-frequency correlations, 
 \begin{align}
 |F_{\omega t}(\omega_s,t_i)|^2&=\left|\smallint d\omega_i
 F_{\omega\omega}(\omega_s,\omega_i)e^{-i\omega_it_i}\right|^2,\\ 
 |F_{t\omega}(t_s,\omega_i)|^2&=\left|\smallint d\omega_s
   F_{\omega\omega}(\omega_s,\omega_i)e^{-i\omega_st_s}\right|^2,
 \end{align}
 which can distinguish between these two cases and break the time-reversal
 ambiguity.

\section{Phase retrieval algorithm}
 The phase retrieval algorithm is shown in Fig.~\ref{fig:mgsa}.
 The algorithm is seeded with an initial guess of the state,
 $F_{\omega\omega}(\omega_s,\omega_i)$, involving a random phase.  In the first
 iteration, we project the state onto the constraint set that satisfies the
 measured intensities in frequency.  This is achieved by replacing the spectral
 amplitudes $|F_{\omega \omega}(\omega_s,\omega_i)|$ with the measured spectral
 amplitudes $\sqrt{I(\omega_s,\omega_i)}$ but keeping the phase, 
 \begin{align}
   F_{\omega \omega}(\omega_s,\omega_i)\rightarrow
   \frac{F_{\omega \omega}(\omega_s,\omega_i)}{|F_{\omega \omega}(\omega_s,\omega_i)|}\sqrt{I(\omega_s,\omega_i)}.
   \label{eq:magconstraint}
 \end{align}
 We then apply the Fast Fourier Transform algorithm (FFT) to obtain an estimate
 of $F_{\omega t}(\omega_s,t_i)$ and again replace the amplitudes $|F_{\omega
 t}(\omega_s,t_i)|$ with the measured amplitudes $\sqrt{I(\omega_s,t_i)}$. This
 is repeated two more times, as in Fig.~\ref{fig:mgsa}, completing one
 iteration of the algorithm. At each iteration, we evaluate the FROG-trace
 error \cite{trebino_frequency-resolved_2012} between the measured and the
 reconstructed joint spectral intensities, which corresponds to the average
 percentage error in each point $(\omega_s, \omega_i)$. An important feature of
 these types of algorithms is that the measured error will always decrease or
 remain constant at each iteration, and will not
 diverge~\cite{gerchberg_practical_1972,saxton_computer_2013}.

 \begin{figure}[t!]
   \centering
   \includegraphics{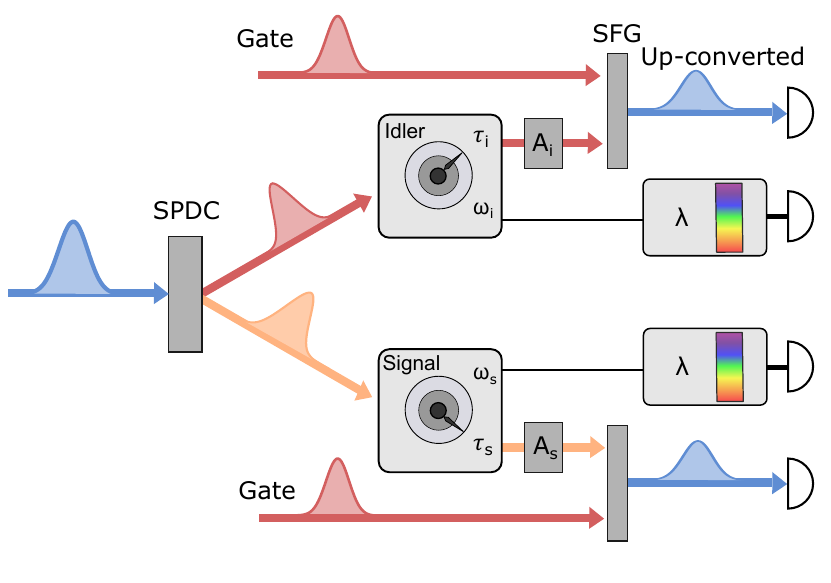}
   \caption{Experimental setup for two-photon state reconstruction.
     Energy-time entangled photons are produced through spontaneous parametric
     downconversion (SPDC). Each photon can be measured in frequency using a
     scanning monochromator or in time by optically gating the single photon
     using sum-frequency generation (SFG) in a nonlinear medium with a strong
     gate pulse.  The delays $\tau_s$ and $\tau_i$ are between the gate pulse
     and the photon on the signal and idler side, respectively. The quadratic
     spectral phases, $A_s$ on the signal photon and $A_i$ on the idler photon,
     are controlled using a fibre and grating compressor on each side.
     Measurements in coincidence of all four combinations of the frequency and
     time of arrival of the photons allow the reconstruction of the joint
     spectral amplitude function using a phase retrieval algorithm.}
   \label{fig:setup}
 \end{figure}

\section{Experiment}
 The setup is schematically depicted in
 Fig.~\ref{fig:setup} and described in detail in
 Refs.~\cite{maclean_direct_2018, maclean_ultrafast_2018}.  Ti:sapphire laser
 pulses [80~MHz, 775~nm, 3.8~W average power, 0.130~ps (s.d.) pulse width], are
 frequency doubled in 2 mm of $\beta$-bismuth borate (BiBO). After spectral
 filtering with a 0.2~nm FHWM bandpass filter, the second harmonic pumps a 5~mm
 BiBO crystal for type-I SPDC generating  energy-time entangled photons at
 823~nm and 732~nm.  These are coupled in single-mode fibres allowing for
 direct, spectrally resolved, or temporally resolved measurements. 
 Spectral measurement are performed via monochromators with a resolution of
 0.1~nm.  Temporal measurements are implemented via optical gating, i.e., via
 noncollinear sum-frequency generation (SFG) with femtosecond laser pulses in
 1~mm of bismuth borate (BiBO) crystal. The electric field of the gate pulse is
 characterized using an SHG-FROG measurement, and we find an intensity pulse
 width of 130~fs (s.d.). 
 Since the experimentally measured intensity in frequency (time) is convolution
 of the joint spectral (temporal) intensity and the filter function of the
 monochromater (temporal gate), the measured data must first be
 deconvolved before it can be used in the phase-retrieval algorithm.  Numerical
 deconvolutions for each intensity measurement are performed using a Wiener
 filter~\cite{press_numerical_2007}, and the resulting output provides the
 intensity distributions, $I(\omega_s,\omega_i), I(t_s,\omega_i),
 I(\omega_s,t_i),I(t_s,t_i)$, used in the algorithm in Fig.~\ref{fig:mgsa}. See
 Appendix~\ref{sec:algorithm} for further details.

 The spectral phase on the photons, 
 \begin{align}
 \phi(\omega_s,\omega_i)\approx
 A_s\left(\omega_s-\omega_{s0}\right)^2+A_i\left(\omega_i-\omega_{i0}\right)^2,
 \end{align}
 is controlled with a combination of normally dispersive single-mode fibre and
 adjustable grating compressor for anomalous
 dispersion~\cite{maclean_direct_2018}, where $A_s$ and $A_i$ are the chirp
 parameters for the signal and idler, respectively. The relative position of
 the gratings inside the compressor sets the magnitude and sign of the overall
 dispersion.  We calibrate both grating compressors using XFROG
 (Cross-correlation FROG) spectrogram measurements between the strong gate
 pulse and a weak laser pulse. The weak laser pulse has the same center
 wavelength and path through the fibre-compressor system as the photons on each
 side. The phase at each relative grating separation is reconstructed using the
 Principal Component Generalized Projection (PCGP) FROG
 algorithm~\cite{kane_recent_1999,trebino_frequency-resolved_2012}.  We find a
 quadratic phase that depends linearly on the grating separation with slopes of
 $(-1360\pm60)$~fs$^2$/mm and $(-2190\pm70)$~fs$^2$/mm for the signal and
 idler, respectively. The difference between the two is attributed to the cubic
 dependence on wavelength of dispersion in a grating
 compressor~\cite{treacy_optical_1969}.

\section{Phase reconstructions}
\begin{figure}[t]
     \centering
     \includegraphics{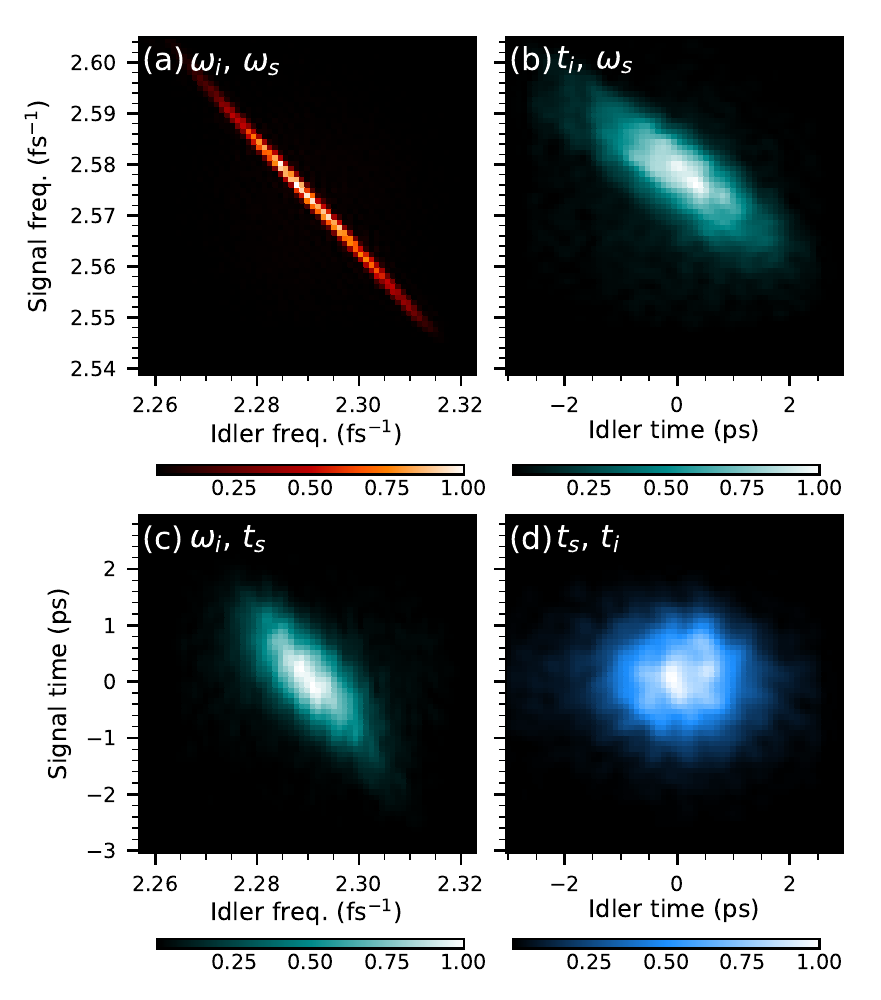}
     \caption{Example deconvolved measured data for two-photon state
       reconstruction when negative dispersion is introduced to signal and
       idler photons.  Combinations of spectral and temporal measurements are
       made in coincidence to obtain the (a) joint spectral intensity, (d)
       joint temporal intensity, and (b,c) correlations between the time and
       frequency of the photon pair for an SPDC state.  We observe strong
       anti-correlations between the measured quantities in (a), (b), (c) and
       very little correlations in (d), indicating the presence of negative
       dispersion on both photons.  After post-processing, the measured
       intensities are used as data constraints for the phase-retrieval
     algorithm.}
    \label{fig:example_data}
  \end{figure}

 We compare the phase retrieval algorithm on measured data for two-photon
 states with different amounts of dispersion. We set the grating compressors on
 the signal and idler side to study four cases: no additional dispersion, with
 extra positive dispersion applied to the idler, with extra negative dispersion
 applied to the signal, and with extra negative dispersion applied on both
 sides.  For the case of a two-photon energy-time entangled state with negative
 dispersion applied to both photons, an example of the four combinations of
 time and frequency measurements is shown in Fig.~\ref{fig:example_data}.
 Background subtraction, a Wiener Filter, and low-pass filters are applied in
 Fig.~\ref{fig:example_data} and prior to the
 reconstruction~\cite{fittinghoff_noise_1995}.  We observe
 strong anti-correlations in the joint spectral intensity
 [Fig.~\ref{fig:example_data}(a)], however, the joint temporal intensity
 [Fig.~\ref{fig:example_data}(d)] is uncorrelated due to the presence of
 dispersion on both photons. The observed shears in both the time-frequency
 intensity plots [Fig.~\ref{fig:example_data}(b-c)] also illustrate the
 presence of negative dispersion.  
 
\begin{figure}[t]
  \centering
  \includegraphics{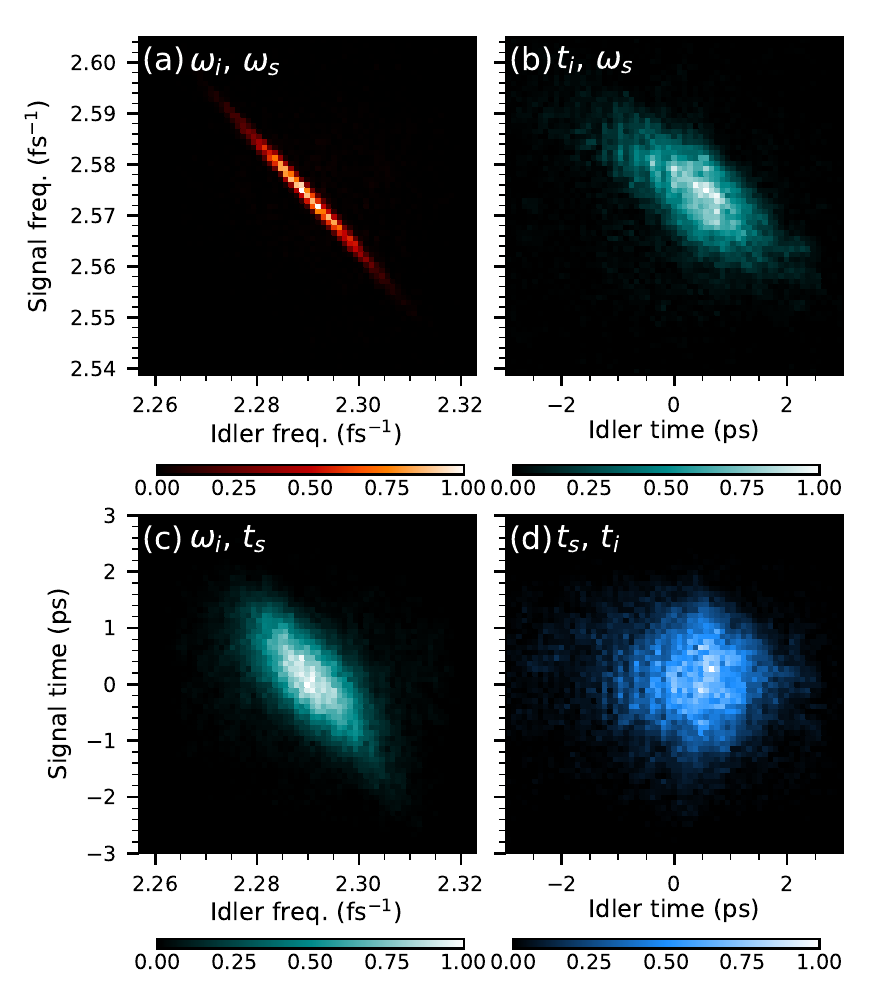}
  \caption{Two-photon state reconstruction. 
    Reconstructed distributions for the (a) the joint spectral intensity, (d)
    the joint temporal intensity, as well the (b,c) time-frequency
    correlations of the measured state in Fig.~\ref{fig:example_data}
    after 1000 iterations of the phase retrieval algorithm. 
    } 
  \label{fig:example_reconstruction}
 \end{figure}

 We map these intensity constraints onto a
 64x64 array and input them into the phase retrieval algorithm, which is run
 for 1000 iterations, a number found heuristically after which no reduction in
 the FROG-trace error is observed.  The intensity of the reconstructed
 wavefunction in frequency and time are shown in
 Fig.~\ref{fig:example_reconstruction}.  The reconstructed intensities are
 compared to the measured data from Fig.~\ref{fig:example_data}(a) and
 Fig.~\ref{fig:example_data}(d).  We find a FROG-trace error between the
 post-processed and reconstructed spectral intensities after 1000 iterations to
 be $(3.64 \pm 0.07)\%$ for the joint spectral intensity and $(7.01 \pm
 0.35)\%$ for the joint temporal intensity. 

 Note that the marginal bandwidths of the joint spectral intensity in the
 reconstruction [Fig.~\ref{fig:example_reconstruction}(a)] are shorter than in
 the original data [Fig.~\ref{fig:example_data}(a)]. Numerical simulations
 suggest that this arises as a result of the phase-matching bandwidth in the
 optical gating.  The effect of the phase mismatch on the reconstruction of
 two-photon states with optical gating is modelled in
 Appendix~\ref{sec:measurement_with_phasematching}.

 \begin{figure*}[ht]
  \centering
  \includegraphics{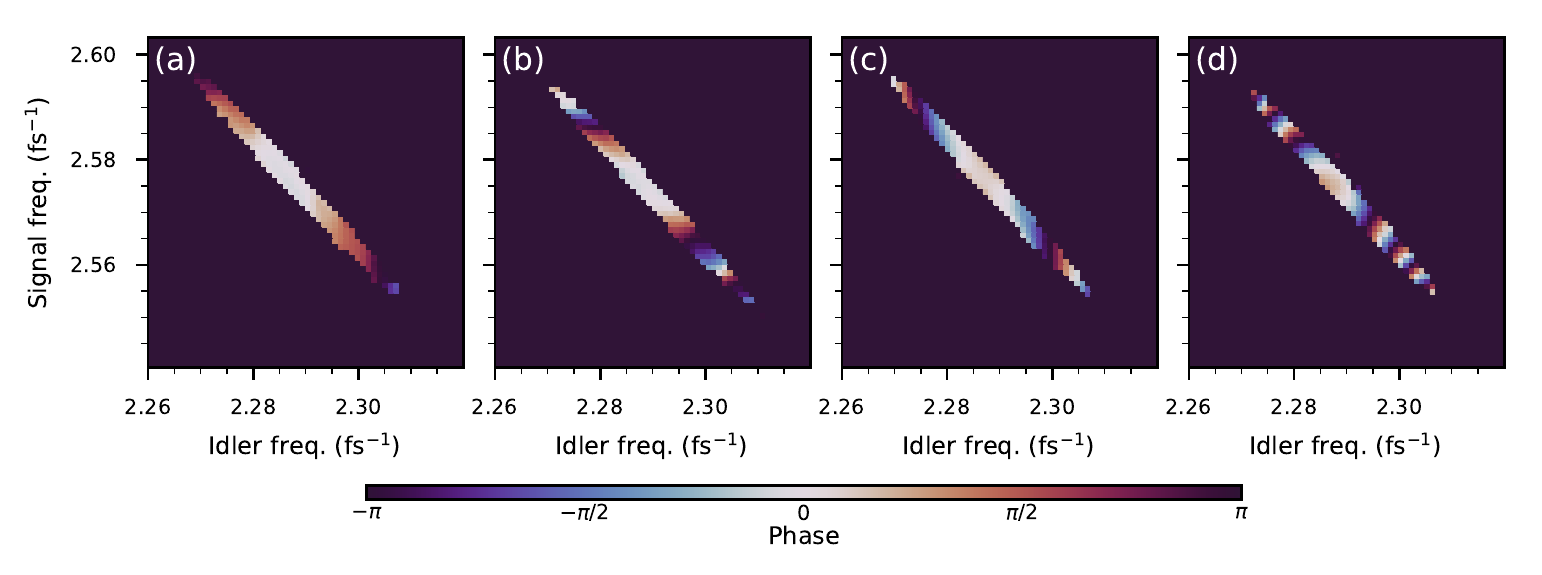}
  \caption{Phase reconstruction of energy-time entangled states.  Reconstructed
    joint spectral phase for energy-time entangled photon pairs with (a) no
    added dispersion, (b) positive dispersion on the signal, (c) negative
    dispersion on the idler, (d) negative dispersion on both the signal and
    idler.  Phase points outside the 2$\sigma$ intensity contours are removed for
    clarity. We observe (a) a relatively flat phase variation, (b) a positive
    quadratic phase variation along the signal axis, (c) a  negative quadratic
    phase variation along the idler axis, (d) and  a negative quadratic phase
    variation along both axes.} 
  \label{fig:phase_reconstruction}
 \end{figure*} 

 Fig.~\ref{fig:phase_reconstruction} shows the reconstructed joint spectral
 phase for the four different cases.  Starting with the case where we attempted
 to minimize the unbalanced dispersion
 [Fig.~\ref{fig:phase_reconstruction}(a)], we observe a relatively flat
 spectral phase.  In this configuration, we measure the time-bandwidth product
 as in Ref.~\cite{maclean_direct_2018} and find
 $\Delta(\omega_s+\omega_i)\Delta(t_s-t_i)=(1.711 \pm 0.005)(0.196 \pm
 0.004)=0.348 \pm 0.006$, verifying the presence of energy-time entanglement.

 For the three cases where dispersion is applied, we fit the reconstructed
 quadratic spectral phase in Fig.~\ref{fig:phase_reconstruction}(b-d). For
 each, we unwrap the 2D phase and perform a polynomial fit to the phase
 distribution.  The corresponding uncertainties are obtained from the variance
 in the fitted spectral phase after performing Monte Carlo simulations assuming
 Poissonian noise.  
 
 When we apply $A_s=(0.026 \pm 0.002)$ ps$^2$ of dispersion on the signal
 photon [Fig.~\ref{fig:phase_reconstruction}(b)], we reconstruct a quadratic
 spectral phase on the signal of $A_s=(0.024 \pm 0.003)$~ps$^2$, observing a
 positive quadratic variation in the phase along the signal (y) axis, modulo
 2$\pi$, with little variations along the idler (x) axis. When we apply
 $A_i=(-0.025 \pm 0.002)$~ps$^2$ of dispersion to the idler photon
 [Fig.~\ref{fig:phase_reconstruction}(c)], we reconstruct a quadratic phase on
 the idler of $A_i=(-0.026 \pm 0.003)$~ps$^2$, observing a negative quadratic
 variation in the spectral phase along the idler (x) axis, with again little
 variations along the signal (y) axis. When we apply $A_s=(-0.036 \pm
 0.003)$~ps$^2$ and $A_i=(-0.043 \pm 0.002)$~ps$^2$ of dispersion to the signal
 and idler [Fig.~\ref{fig:phase_reconstruction}(d)], we reconstruct a quadratic
 phase on the signal and idler of $A_s=(-0.036 \pm 0.004)$~ps$^2$ and
 $A_i=(-0.028 \pm 0.003)$~ps$^2$, respectively. For this case, we observe a
 negative quadratic variation along the diagonal x-y axis.
 
 When dispersion is applied to only one photon,
 Fig.~\ref{fig:phase_reconstruction}(b) and
 Fig.~\ref{fig:phase_reconstruction}(c), the phase obtained using the
 phase-retrieval algorithm corresponds to the reconstructed phases measured
 using the XFROG algorithm. In the last case,
 Fig.~\ref{fig:phase_reconstruction}(d), we find a discrepancy between the two.
 This, again, is likely due to the effect of the phase mismatch on the temporal
 measurements and on the subsequent reconstruction of two-photon states, which
 will be more pronounced for the photons which have much larger bandwidth than
 for the weak pulse used for the XFROG reconstructions (see
 Appendix~\ref{sec:measurement_with_phasematching}).

\section{Conclusion} 
 We have demonstrated a method to recover ultrafast two-photon energy-time
 entangled pulses.  Our technique is based on a method of alternate projections
 that iterates between the frequency and time domains imposing the measured
 intensity constraints at each iteration.  The use of nonlinear phenomena,
 i.e., optical gating, to measure the timing correlations is an artifact of the
 time scales at play and is not a fundamental requirement. For sufficiently
 long pulses, there may exist photodetectors that can measure the temporal
 intensity directly~\cite{korzh_demonstrating_2018}.  For subpicosecond
 resolution involving optical gating, the effect of phase-matching in the
 upconversion could be reduced using shorter crystals or
 angle-dithering~\cite{oshea_increased-bandwidth_2000}.  

 In our simulations, the reconstruction fidelity seems to depend on the amount
 of entanglement in the initial state, and uncovering the reason for this is
 the subject of future work.
 Moreover, extensions of this algorithm to characterize two-photon mixed states
 may be possible based on techniques used to reconstruct partially coherent
 light~\cite{thibault_reconstructing_2013,bourassin-bouchet_partially_2015},
 removing assumptions about the purity of the quantum states.  Measurement and
 reconstruction capabilities similar to those available in ultrafast optics
 will be essential for developing new applications in quantum state engineering
 and ultrafast shaping of entangled photons, paving the way to characterizing
 and manipulating high-dimensional quantum states of light. 

 The authors thank J.M. Donohue and F. Miatto for fruitful discussions.  This
 research was supported in part by the Natural Sciences and Engineering
 Research Council of Canada (NSERC), Canada Research Chairs, Industry Canada
 and the Canada Foundation for Innovation (CFI).

\appendix
\section{Phase retrieval algorithm}
\label{sec:algorithm}

 The algorithm to reconstruct the phase of energy-time entangled states is
 divided into two main parts: post-processing and phase retrieval. 
 In the first part, post-processing, for each experimentally measured
 time-frequency correlation plot, the data is interpolated onto a 2D square grid, $H(x,y)$,
 of size 64x64, where, here, we use $x,y$ to represent either the measured time
 or frequency variables.  We apply background subtraction using a corner
 suppression routine~\cite{trebino_frequency-resolved_2012}. The numerical
 deconvolution is performed using a Wiener filter,
 \begin{align}
 W\left( {{k_x},{k_y}} \right) = \frac{{G{{\left( {{k_x},{k_y}} \right)}^ *
 }}}{{{{\left| {G\left( {{k_x},{k_y}} \right)} \right|}^2} + \alpha }},
 \end{align}
 where $G(k_x,k_y)$ is the filter function which we obtain by taking the
 Fourier transform of the instrument response functions. These are approximated
 as Gaussian functions with the instrument resolutions obtained experimentally
 for the spectral (0.1~nm) and temporal (0.130~ps) measurements.  The parameter
 $\alpha$ takes into account the amount of noise in the system and will
 typically depend on $k_x$ and $k_y$.  Here, we approximate it as a
 constant, $0.05\le\alpha\le0.2$, which is obtained heuristically for each
 reconstruction.  A low-pass filter is also applied by multiplying the
 Wiener filter $W\left( k_x,k_y\right)$  by a top-hat function $T(k_x,k_y)$ of radius,
 $\varrho N/2$ in pixels, and setting all the values outside $\varrho N/2$ to
 0, where $N$ is the size of the grid, and $0.8\le\varrho\le1$. The
 deconvolved intensities are then obtained with the inverse Fourier
 transform, 
 \begin{align}
   I\left( {x,y} \right) = FT^{-1}\left[ {H\left( {{k_x},{k_y}}
   \right)W({k_x},{k_y}) T(k_x,k_y)} \right], 
   \label{}
 \end{align}
 where $H(k_x,k_y)$ is the Fourier Transform of the experimentally measured intensities
 $H(x,y)$. The resulting deconvolved intensities $I(x,y)$ are used as the
 physical constraints in the phase retrieval algorithm.

In the second part, the phase retrieval algorithm is seeded with an initial
 guess of the state, which can consist of the measured amplitudes with a random
 phase. Steps (1-8) are used when all four intensity constraints are applied. 
 \begin{enumerate}
   \item Replace the magnitude of $F_{\omega\omega}(\omega_s,\omega_i)$ with the measured values     
	\begin{equation}
	F_{\omega\omega}^\prime(\omega_s,\omega_i)=\frac{F_{\omega\omega}(\omega_s,\omega_i)}{|F_{\omega\omega}(\omega_s,\omega_i)|}\sqrt{I(\omega_s,\omega_i)}.
	\end{equation}	     
   \item Evaluate the inverse Fourier transform $F_{\omega\omega}(\omega_s,\omega_i)$ to obtain an estimate of $F_{\omega t}(\omega_s,t_i)$.
   \item Replace the magnitude of $F_{\omega t}(\omega_s,t_i)$ with the measured values
   \begin{equation}
   F_{\omega t}^\prime(\omega_s,t_i)=\frac{F_{\omega t}(\omega_s,t_i)}{|F_{\omega t}(\omega_s,t_i)|}\sqrt{I(\omega_s,t_i)}.
   \end{equation}
   \item Evaluate the inverse Fourier transform $F_{\omega t}^\prime(\omega_s,t_i)$ to obtain an estimate
     of $F_{tt}(t_s,t_i)$.
   \item Replace the magnitude of $F_{tt}(t_s,t_i)$ with the measured values
   \begin{equation}
   F_{tt}^\prime
     (t_s,t_i)=\frac{F_{tt}(t_s,t_i)}{|F_{tt}(t_s,t_i)|}\sqrt{I(t_s,t_i)}.
   \end{equation}
   \item Evaluate the Fourier transform $F_{tt}^\prime(t_s,t_i)$ to obtain an estimate of
     $F_{t \omega}(t_s,\omega_i)$.
   \item Replace the magnitude of $F_{t\omega}(t_s,\omega_i)$ with the measured values
   \begin{equation}
   F_{t\omega}^\prime
     (t_s,\omega_i)=\frac{F_{t\omega}(t_s,\omega_i)}{|F_{t\omega}(t_s,\omega_i)|}\sqrt{I(t_s,\omega_i)}.
   \end{equation}
   \item Evaluate the Fourier transform $F_{\omega t}^\prime(\omega_s,t_i)$ to obtain an estimate of
     $F_{\omega\omega}(\omega_s,\omega_i)$.
 \end{enumerate}
 
 The time to run the algorithm depends on the size of the arrays being used and
 the number of iterations. For the case in Fig. 4, using a 64x64 array,
 the entire procedure, including loading the data, applying filtering and
 deconvolution algorithms, and running the phase retrieval algorithm for 1000
 iterations takes about $(10 \pm 4)$s, averaged over 100 runs and using a
 laptop computer (i7-4650U CPU @2.3GHz with 8 Gb of RAM).\\     
 
\section{Reconstructing two-photon states with optical gating}
\label{sec:measurement_with_phasematching}

 To test the numerical deconvolution and the phase-reconstruction algorithm
 described above using realistic temporal measurements, we construct a
 numerical model of optical gating with thick crystals and consider its effect
 on the measurement and reconstruction of energy-time entangled two-photon
 states.  

 An energy-time entangled two-photon state as in Eq.~\ref{eq:psi} with a
 joint-spectral amplitude for the signal $\omega_s$
 and idler $\omega_i$ frequencies is modelled with a two-dimensional correlated
 Gaussian function,
 \begin{widetext}
 \begin{align}
   \begin{split}
   F_{\omega \omega}&\left( \omega_s,\omega_i \right) = 
   \frac{1}{\sqrt{2\pi\sigma_{\omega_s}\sigma_{\omega_i} } \left( 1 - \rho_{\omega}^2 \right)^{1/4}} \\
   &\exp \left( { -\frac{1}{{2\left( {1 - {\rho_{\omega} ^2}} \right)}}\left[ {\frac{{{{\left(
     {{\omega _s} - {\omega _{s0}}} \right)}^2}}}{{2{\rm{\sigma }}_{\omega_s}^2}} 
     + \frac{{{{\left( {{\omega _i} - {\omega _{i0}}} \right)}^2}}}{{2\sigma
       _{\omega_i}^2}} 
 - \frac{\rho_{\omega} \left(\omega_s-\omega_{s0}\right)\left(\omega _i-\omega
 _{i0}\right)}{\sigma_{\omega_s}\sigma_{\omega_i}}} \right]}
 \right).
 \end{split}
 \label{eq:Fgauss}
  \end{align}
  \end{widetext}

 The marginal frequency bandwidths, $\sigma_{\omega_i}$ and
 $\sigma_{\omega_s}$, are set to the values measured experimentally. 
 The correlation parameter $\rho_{\omega}=\Delta(\omega_s\omega_i)/\Delta
 \omega_s \Delta \omega_i$ describes the statistical correlations between the
 frequency of the signal and idler modes and is related to the purity of the
 partial trace, $P=\sqrt{1-\rho_{\omega}^2}$. When $\rho_{\omega}=0$, the
 joint-spectral amplitude $F(\omega_s,\omega_i)$ factorizes and the state is
 separable, whereas when $\rho_{\omega}\rightarrow-1$, the photons are
 perfectly anticorrelated in frequency. Furthermore, when the marginal
 bandwidths are equal, $\sigma_{\omega_s}=\sigma_{\omega_i}$, the
 time-bandwidth product for the Gaussian state in Eq.~\ref{eq:Fgauss}  is
 $\Delta(\omega_s+\omega_i)\Delta(t_s-t_i)=\sqrt{(1+\rho)/(1-\rho)}$.
 We apply a quadratic phase to the state, 
 \begin{align}
   F_{\omega\omega}(\omega_s,\omega_i)\rightarrow
   F_{\omega\omega}(\omega_s,\omega_i)e^{iA_s\left(\omega_s-\omega_{s0}\right)^2+iA_i\left(\omega_i-\omega_{i0}\right)^2},
   \label{eq:Fphase}
 \end{align}
 where $A_s$ and $A_i$ are the chirp parameters on the signal and idler,
 respectively. 

 In the SFG process used for optical gating, a photon and a strong gate pulse
 in the near-infrared (NIR) are up-converted to produce a higher energy photon
 in the ultraviolet.  If the photons are dispersed before the optical gating,
 high and low frequencies components will arrive at different times in the
 nonlinear medium. In the presence of phase mismatch, the upconverted
 frequencies associated to these high and low frequency components can lie
 outside the phase-matching bandwidth of the crystal, and consequently will be
 suppressed.  As a result,  phase mismatch in optical gating changes the
 measured intensity correlations and therefore changes the deconvolved
 intensity constraints that are applied in the phase retrieval algorithm.  
 
 To account for this effect, we model the optical gating as sum-frequency
 generation process in the low-efficiency regime between the photons on each
 side and a gate pulse with centre frequency $\omega_g$ and a pulse duration of
 0.130~ps, leading to upconverted frequencies $\omega_{u_s}=\omega_s+\omega_g$
 and $\omega_{u_i}=\omega_i+\omega_g$ on the signal and idler side,
 respectively.  The gate pulse is modeled with a Gaussian temporal profile,
 \begin{align}
 \begin{split}
 &{G}\left( {{\omega _g},{\tau_g}} \right)= 
 \frac{1}{{{{\left( {2\pi {\rm{\sigma }}_g^2} \right)}^{\frac{1}{4}}}\;}}\;
 e^{{ - \frac{{{{\left( {{\omega _g} - {\omega _{g0}}}
 \right)}^2}}}{{4{\rm{\sigma }}_g^2}}} +i\tau_g\left( \omega_g -
 \omega_{g_0}\right)},
 \label{eq:gate_pulse}
 \end{split}
 \end{align}
 with marginal bandwidth $\sigma_g$, and delay $\tau_g$.  
 For the purpose of this simulation, we assume the spectral measurements have
 high resolution such that they can be represented by delta functions,
 \begin{align}
   H_{\omega\omega}(\omega_s,\omega_i)\approx|F(\omega_s,\omega_i)|^2 ,
   \label{eq:Hww}
 \end{align}
 and the three intensity measurements involving optical gating are calculated
 via the following,
 \begin{widetext}    
 \begin{align}
   &H_{\tau\omega}\left( {{\tau _s},{\omega _i}} \right) 
  = \int d{\omega _{u_s}}{\left| {\int d{\omega _s}G\left( {{\omega _{u_s}} - {\omega _s},{\tau _s}} \right)
      {\rm{\Phi_{\rm{SFG}} }}\left( {{\omega _s},{\omega _{u_s}} - {\omega _s},{\omega
  _{u_s}}} \right)\;F_{\omega \omega}\left( {{\omega _s},{\omega _i}} \right)} \right|^2},
  \label{eq:Htw}
   \\
   &H_{\omega\tau}\left( {{\omega_s},{\tau _i}} \right) 
  = \int d{\omega _{u_i}}{\left| {\int d{\omega _i}
      G\left( {{\omega _{u_i}} - {\omega _i},{\tau _i}} \right) 
      {\rm{\Phi_{\rm{SFG}} }}\left( {{\omega _i},{\omega _{u_i}} - {\omega
  _i},{\omega _{u_i}}} \right)\; F_{\omega \omega}\left( {{\omega _s},{\omega _i}} \right)}
\right|^2},
   \label{eq:Hwt}
   \\
   \label{eq:Htt}
   &H_{\tau\tau}\left( \tau_s,\tau_i \right) 
  = \int d\omega_{u_s}d\omega_{u_i} \Bigg|\int d\omega_s d\omega_i
  G\left(\omega_{u_s}-\omega_s,\tau_s\right) 
  \Phi_{\rm{SFG}}(\omega _s,\omega_{u_s} - \omega_s,\omega_{u_s})\\
  &\hspace{4.7cm}\nonumber
  \times~ G\left(\omega_{u_i}-\omega_i,\tau_i\right) 
  \Phi_{\rm{SFG}}(\omega _i,\omega_{u_i} - \omega_i,\omega_{u_i})
  F_{\omega \omega}(\omega_s,\omega_i) \Bigg|^2,
 \end{align}
 \end{widetext}
 where the gate pulse $G$ is the same on both sides but with delays $\tau_i$
 and $\tau_s$ introduced. The phase matching function is
 \begin{align}
   \begin{split}
 \Phi_{\rm{SFG}}(\omega_j,\omega_{u_j} &- \omega_j,\omega_{u_j})=
 e^{-i\frac{\Delta k L}{2}}\rm{sinc}\left(\frac{\Delta k L}{2}\right),
 \end{split}
   \label{eq:sinc}
 \end{align}
 where the phase-mismatch, 
 \begin{align}
   \begin{split}
   \Delta k(\omega_j,\omega_{u_j} - \omega_j,\omega_{u_j})=
   &\frac{n_e(\omega_j)\omega_j}{c}+\frac{n_o(\omega_{u_j})\omega_{u_j}}{c} \\
   &+\frac{n_e(\omega_{u_j} - \omega_j)(\omega_{u_j} - \omega_j)}{c},
 \end{split}
   \label{eq:Deltak}
 \end{align}
 is calculated for type-I SFG with different crystal lengths $L$ and the
 experimentally measured wavelengths. The phase-matching bandwidth can be
 estimated from the range of frequencies contained in $\Delta k L=\pi$.
 Upconverted frequencies outside this range are suppressed.  All integrals are
 evaluated numerically. 

 We model all the steps in the phase-retrieval process.  We numerically create
 frequency anti-correlated states using Eq.~\ref{eq:Fgauss} and
 Eq.~\ref{eq:Fphase}, with the same centre wavelength and bandwidth as those
 measured experimentally, but with different amounts of applied spectral
 phases, given by the chirp parameters $A_s$ and $A_i$.  We calculate the four
 joint correlations in frequency and time with Eqs.~(\ref{eq:Hww}-\ref{eq:Htt})
 using different lengths of BiBO for optical gating, apply the numerical
 deconvolution to each intensity measurement as described in
 Appendix~\ref{sec:algorithm}, and insert these as constraints for the phase
 retrieval algorithm.  After reconstruction, we unwrap the spectral phase of
 the reconstructed joint spectral amplitude function and fit it to a
 third-order two-dimensional polynomial.  

 \begin{figure*}[th!]
   \centering
   \includegraphics[scale=1.0]{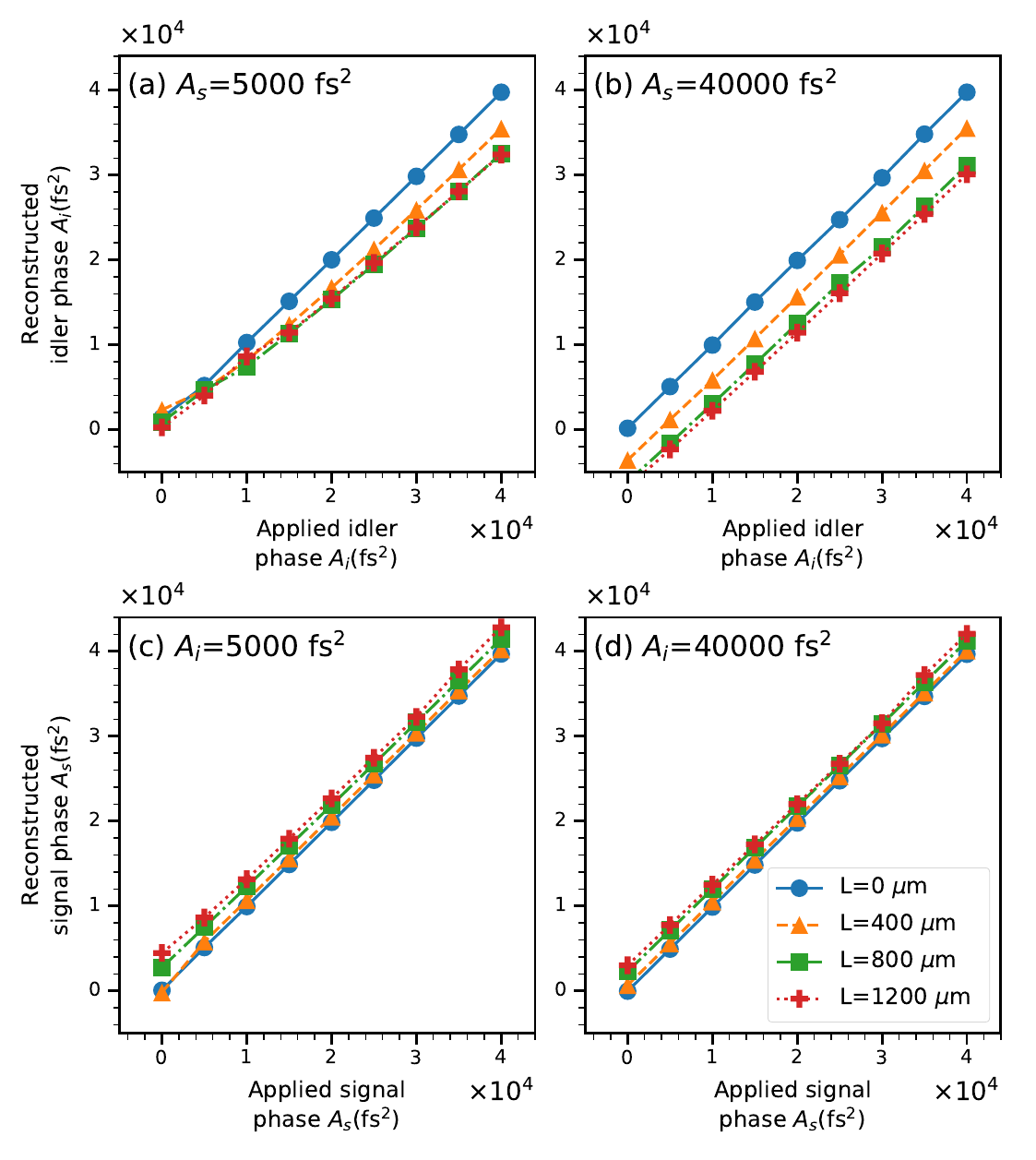}
   \caption{Effect of phase mismatch on the reconstructed spectral phase. 
      We model the effect of optical gating with different lengths $L$ of BiBO
      on the reconstructed phase. The reconstructed phase is compared to the
      applied phase for four different cases. The signal chirp parameter is
      fixed to the values of (a) $A_s=5,000$ fs$^2$ and (b) $A_s=40,000$ fs$^2$
      while the idler chirp parameter $A_i$ is varied. The idler chirp
      parameter is fixed to the values of (c) $A_i=5,000$ fs$^2$ and (d)
      $A_i=40,000$ fs$^2$ while the signal chirp parameter $A_s$ is varied. At
      $L=0$ $\mu$m, the reconstructed phase is the same as the applied phase.  As
      $L$ is increased, phase mismatch becomes more important and this changes the
      value of the reconstructed phase.}
   \label{fig:model_reconstruction}
 \end{figure*}

 The reconstructed spectral phases are compared to the applied spectral phases
 in Fig.~\ref{fig:model_reconstruction} for different lengths of BiBO used in
 optical gating and for different applied spectral phases. In
 Fig.~\ref{fig:model_reconstruction}(a) and
 Fig.~\ref{fig:model_reconstruction}(b), the signal chirp parameter $A_s$ is
 kept fixed while the idler chirp parameter is varied, whereas in
 Fig.~\ref{fig:model_reconstruction}(c) and
 Fig.~\ref{fig:model_reconstruction}(d), the idler chirp parameter $A_i$ is
 kept fixed while the signal chirp parameter $A_s$ is varied.  When the length
 of the crystal is set to zero ($L=0$ $\mu$m), the reconstructed phase
 corresponds exactly to the applied phase, and the line at $L=0$~$\mu$m appears
 at 45 degrees with a slope of one. As the length of the crystal increases, we
 find that the slope remains fairly constant at 45 degrees, but the offset
 depends on the configuration. For example, comparing
 Fig.~\ref{fig:model_reconstruction}(a) and
 Fig.~\ref{fig:model_reconstruction}(b), we find the values of the
 reconstructed idler chirp parameter $A_i$ depend on whether the signal chirp
 parameter has a value of $A_s=5,000$ fs$^2$
 [Fig.~\ref{fig:model_reconstruction}(a)] or $A_s=40,000$ fs$^2$
 [Fig.~\ref{fig:model_reconstruction}(b)].  The difference between the
 reconstructed and applied phase in Fig.~\ref{fig:model_reconstruction} also
 becomes larger for longer crystals where the phase matching function is more
 restrictive.

 \bibliography{quantum_FROG}

\end{document}